\documentstyle[11pt,epsf]{article}
\begin{document}

\begin{center}
 J. Phys. A: Math. Gen. {\bf 31} (6 Nov. 1998) 8835-8839 (quant-ph/9709019)
\end{center}

\bigskip

\begin{center}
{\bf Supersymmetric one-parameter  strict isospectrality for the
attractive $\delta$ potentials}

L.J. Boya$^{\dagger}$,
H.C. Rosu$^{\ddagger}$,
A.J. Segu\'{\i}-Santonja$^{\dagger}$,
J. Socorro$^{\ddagger}$,
F.J. Vila$^{\dagger}$

$^{\dagger}$
{\it Departamento de F\'{\i}sica Te\'orica,
Universidad de Zaragoza, 50009 Zaragoza,
Spain}\\
$^{\ddagger}$
{\it Instituto de F\'{\i}sica de la Universidad de Guanajuato, Apdo Postal
E-143, Le\'on, Guanajuato, M\'exico}

\end{center}

\begin{abstract}

The Schr\"odinger equation with attractive $\delta$ potential has been
previously studied in the supersymmetric quantum mechanical approach by a 
number of authors, but they
all used only the particular superpotential solution. Here,
we introduce a one-parameter family of strictly isospectral attractive
$\delta$ function potentials, which is based on the general superpotential
(general Riccati) solution, we study the problem in some detail and suggest
possible applications.

\end{abstract}



The  $\delta (x)$ (pseudo)potential is a well known `zero-range'
potential with applications in solid state physics \cite{Bez}
and many other areas. It has
been used as a textbook example for many mathematical procedures in
quantum mechanics. One such technique, Witten's supersymmetric scheme \cite{W},
has been employed for the attractive delta potential by several authors
\cite{sd1,sd2,sd3}. However, in all those studies there is a missing point,
namely
all the authors so far used only the particular Witten superpotential $W_{0}$,
which is related to the ground state wavefunction in the well-known way
$u_{0}=e^{-\int ^{x}W_{0}}$, and no mention is made of the general
superpotential, i.e., the general Riccati solution for the $\delta$ potential
case. In this work we present the supersymmetric approach
to the attractive delta potential problem based on the general
superpotential.



To help the reader to better understand our problem we start with
its underlying mathematical scheme. Thus, we consider a Riccati equation (RE)
of the type $W^{'}= -W^2+V_{2}(x)$ for which we suppose to know a particular
solution $W_0$. Let $W_{1}=W_{0}+u$ be the second solution. By substituting
$W_1$ in RE one gets the Bernoulli equation $u^{'}=-u^2-2W_{0}u$, which by
means of $u=1/v$ is turned into the first order linear differential equation
$v^{'}-2W_{0}v-1=0$. The latter one can be solved by employing the
integration factor $f_0=e^{-2\int ^{x}W_{0}}$, leading to the solution
$v=f_{0}^{-1}(C+\int^{x}f_{0})$, where $C$ is an arbitrary integration
constant.
Coming back to the general Riccati solution, one gets
$$
W_1=W_0+\frac{f_0}{C+\int^{x}f_0} = W_0+ \frac{d}{dx}\Bigg[
\ln(C+\int^{x}f_{0})\Bigg]~.
\eqno(1)
$$

The point now is that in the process of factorizing the
one-dimensional Schr\"odinger operator $-d^2/dx^2 +V_{1}(x)$
the aforementioned Riccati solutions occur in the non-operatorial part
of the factorization operators as follows.
$W_0$ occurs in the case of
Witten's factorization \cite{W} $(-d/dx +W_0)(d/dx +W_0)
(\equiv A_{0}^{\dagger}A_{0})$,
whereas $W_1$ occurs for Mielnik's factorization \cite{M}
$(-d/dx +W_1)(d/dx +W_1) (\equiv A_{1}^{\dagger}A_{1})$. Notice that
$[A_{0}^{\dagger},A_{0}]=2W_{0}^{'}$, whereas
$[A_{1}^{\dagger},A_{1}]=2W_{1}^{'}$. 
We further notice that $\sqrt{f_{0}}$ is the
ground state (nodeless) wavefunction of $V_1$ and $\Delta V _{0}=-2W_{0}^{'}$
is the Darboux transform contribution to the potential $V_{1}$, leading to a 
new potential $V_{1,D0}=V_{1}-2W_{0}^{'}\equiv V_{2}$, which in
supersymmetric quantum mechanics is
known as the supersymmetric partner of the initial potential $V_{1}$.
Even more interesting is that $\frac{\sqrt{f_0}}{C+\int ^{x}f_{0}}$ can
be interpreted as the ground state wavefunction corresponding to
Mielnik's superpotential (see below),
and $\Delta V_{1}=-2W_{1}^{'}$ can be thought of as
the general Darboux transform part in the potential. Therefore, there is
a one-parameter family of Darboux potentials given by
$V_{1,D1}=V_{1}-2W_{1}^{'}$, which are strictly isospectral to the
initial one, in the sense that each member of the family has the same 
supersymmetric
partner $V_{2}$ and the same energy eigenvalues and scattering amplitudes
as $V_{1}$.
In terms of the ground state wavefunction of $V_1$, $\psi _{0}=\sqrt{f_0}$,
each member of the strictly isospectral family of potentials reads
$$
V_{iso;i}=V_{1}+\Delta V_{1}=V_{1}(x)-2\frac{d^2}{dx^2}\ln \left(C_{i}+
        \int^{x}f_{0}\right)
\eqno(2)
$$
or
$$
V_{iso;i}=V_{1}(x)-\frac{4\psi _{0}
        \psi _{0}^{'}}{C_{i}+\int^{x}\psi _{0}^{2}}
+\frac{2\psi _{0}^{4}}{(C_{i}+\int^{x}\psi _{0}^{2})^2}~.
\eqno(3)
$$
For all half-line potentials the lower limit of the integral term is
zero, whereas for the full-line potentials is $-\infty$.
The ground state wavefunctions of this family are of the type $\psi _{0,iso}=
\frac{\psi _{0}}{(C+\int^{x}\psi _{0}^{2})}$. Indeed, one can write
$$
W_1=
-\frac{d}{dx}
\ln\Bigg[\frac{\psi _{0}}{(C+\int ^{x}\psi _{0}^{2})}\Bigg]
=-\frac{d}{dx}\ln \psi _{0,iso}~,
\eqno(4)
$$
which is the supersymmetric formula introducing the superpotential in terms
of the ground state wavefunction. If one consider these
isospectral functions as quantum mechanical wavefunctions,
the problem of the normalization constant should be contemplated. It is
easy to see that the normalization constant is $N_{iso}=
\sqrt{C(C+1)}$ \cite{rev} and as such $C$ is not allowed to be in $[-1,0]$.
The $C=0$ limit is known as the
Pursey limit \cite{P},
whereas the $C=-1$ limit is the Abraham-Moses limit \cite{AM}. However, in the 
present work we shall consider both the case with the normalization constant 
included and the case without it.



Let us now pass to the attractive $g\delta(x)$ potential, where $g<0$ gives the
strength of the interaction (the binding power). It has been shown
that $W_{0}=\frac{g}{2}
{\rm sign(x)}$ \cite{sd2}. In other words,
$A_{0}=d/dx +\frac{g}{2}{\rm sign(x)}$
and $A^{\dagger}_{0}=-d/dx +\frac{g}{2} {\rm sign(x)}$. Indeed, one cannot
use the Heaviside step function as the superpotential since its square is
not a constant. Therefore, one should work with the sign function, which is
a combination of step functions. $A^{\dagger}\psi _0=0$ implies
$\psi _0=\sqrt{-g/2}e^{g|x|/2}$ and this ground state wavefunction is the only
one of the bound spectrum at the energy $E_0=-g^2/4$. Thus, this state will
be deleted from the spectrum of the partner potential, which is purely
repulsive. However, the situation is by far the more interesting in the case of
the strictly isospectral construction as one can see in the following.

A simple calculation shows that
$$
{\cal I}(x)=\int _{-\infty}^{x}\psi _{0}^{2}(x')dx'=
-\frac{1}{2}{\rm sign}(x)e^{g|x|}+\frac{{\rm sign}(x)}{2}+\frac{1}{2}~.
\eqno(5)
$$
Thus one gets
$$
V_{iso}= g\delta _{iso}(x)=g\delta(x)+
2g^2\frac{{\cal C}{\rm sign}(x)e^{-g|x|}}{(1-{\cal C}{\rm sign(x)}e^{-g|x|})^2}
\eqno(6)
$$
and the isospectral wavefunction reads
$$
\psi _{0,iso}
=-\sqrt{-2g}\sqrt{C(C+1)}
\frac{{\rm sign}(x)e^{-g|x|/2}}{(1-{\cal C}{\rm sign(x)}e^{-g|x|})}~,
\eqno(7)
$$
where ${\cal C}=
2C+{\rm sign}(x)+1$.
The eigenvalue corresponding to the isospectral wavefunction is the same
as for the common delta bound state, i.e., $E_{0}=-g^2/4$.
The analysis of Eqs.~(6) and (7) shows that possible singularities are to be 
found for $C$ in the interval (-1,-1/2], which is excluded when one considers
normalizable isospectral wavefunctions. However for non-normalizable 
solutions these singularities should be taken into account.
The plots we did for the isospectral potentials as a function of the 
isospectral parameter (figure~1) display a
shallow potential well on the negative half-line  
moving toward the origin where it is absorbed by the delta singularity there,
and on the positive half-line a tail dying off at increasing $C$
We also present plots showing the
behaviour of the normalized isospectral wavefunctions for the same values of 
the $C$ parameter as for the potentials (see figure~2).
Moreover, figures~3 and 4 display the moving singularity structure
when we do not introduce the normalization constant in equation~(7). 
In summary, we believe that the strictly isospectral extension of the
attractive
$\delta$ potential introduced here may be relevant for many applications,
once one allows for a physical origin of the $C$-dependence. For example,
the parameter $C$ may express the effect of static and/or moving
distant boundaries, as well as sample-size dependence \cite{B,C}.
If one does not discard as unphysical the non-normalizable isospectral solutions,
one may think of the isospectral method as allowing to introduce singularities 
in both wavefunctions and potentials which apparently are required to explain
the extra losses of ultracold neutrons at the walls \cite{ucn}.


{\bf Acknowledgments}

This work was partially supported by the research grants
AEN96-1670 (CSIC) and ERBCHRX-CT92-0035 and by the
CONACyT projects 458100-5-25844E and 3898P-E9608.


\bigskip
\bigskip


{\bf Figure Captions}






 

 {\bf Figure 1.}

Darboux potential contributions for $C$ equal
to .00001, .10001, 1.10001, 5.10001, for $g=-1$ from left to right.

 {\bf Figure 2.}

The corresponding isospectral wavefunctions for the same values of $C$ as
in figure~1 showing how in the infinite limit of $C$ one recovers the
original $\delta$ wavefunction. Actually, for rather low values of $C$, the 
isospectral $\delta$ wavefunction is already very close in shape to the 
original one.

 {\bf Figure 3.}

Darboux potential contributions for $C$ equal to -1.4, -0.9, (up), and,
-0.6, -0.3 (down) for $g=-1$.

 {\bf Figure 4.}

Non-normalizable isospectral wavefunctions for the same values of $C$ as 
in figure~3, together with the original ground state $\delta$ wavefunction
displayed in the first plot of the figure ($g=-1$).


\newpage



\vskip 2ex
\centerline{
\epsfxsize=280pt
\epsfbox{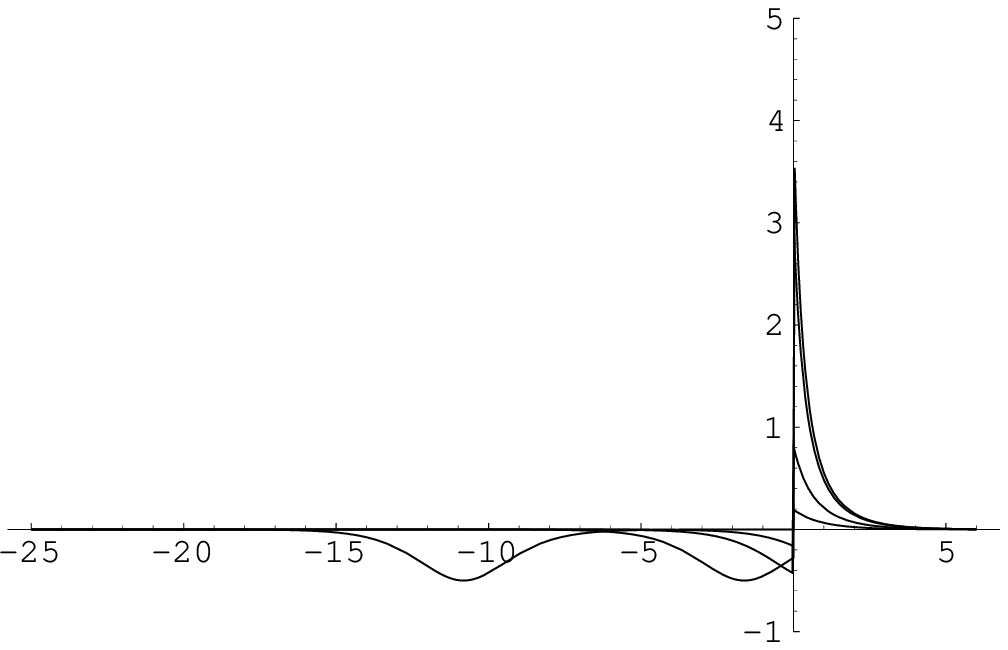}}
\vskip 4ex
\begin{center}
{\small{Figure 1}\\
}
\end{center}

\vskip 2ex
\centerline{
\epsfxsize=280pt
\epsfbox{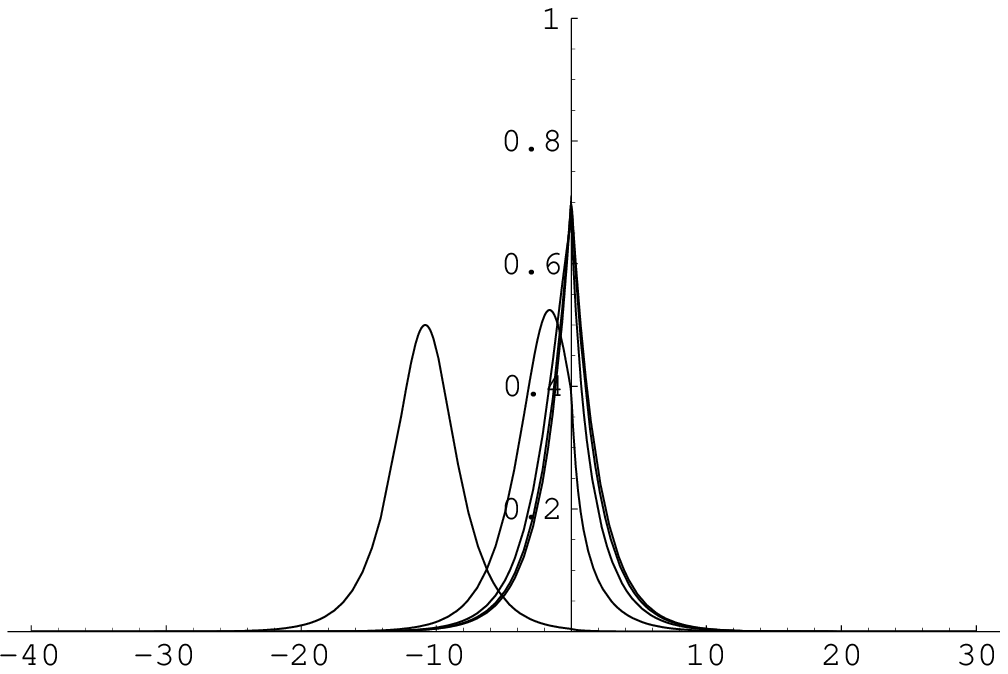}}
\vskip 4ex
\begin{center}
{\small{Figure 2}\\
}
\end{center}

\vskip 2ex
\centerline{
\epsfxsize=280pt
\epsfbox{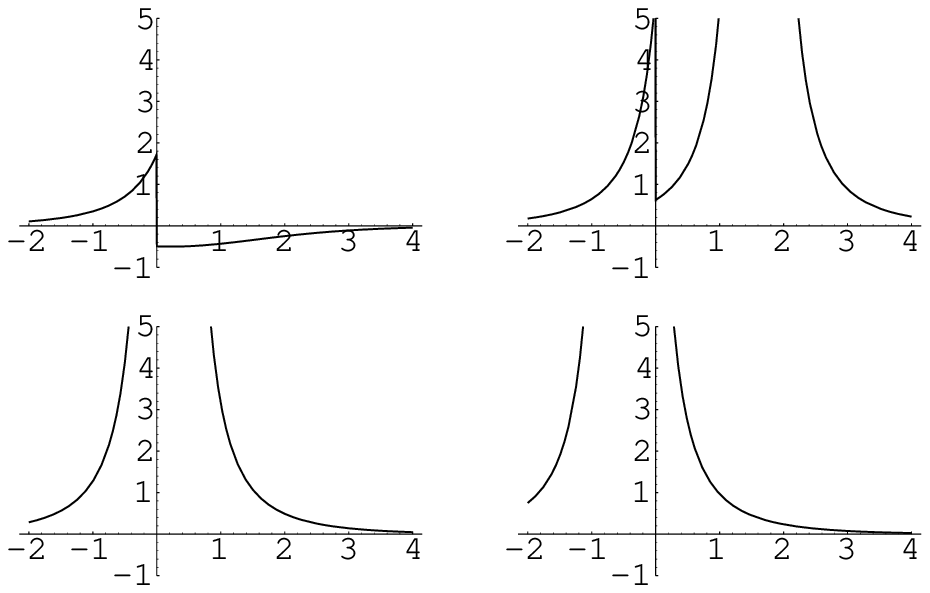}}
\vskip 4ex
\begin{center}
{\small{Figure 3}\\
}
\end{center}

\vskip 2ex
\centerline{
\epsfxsize=280pt
\epsfbox{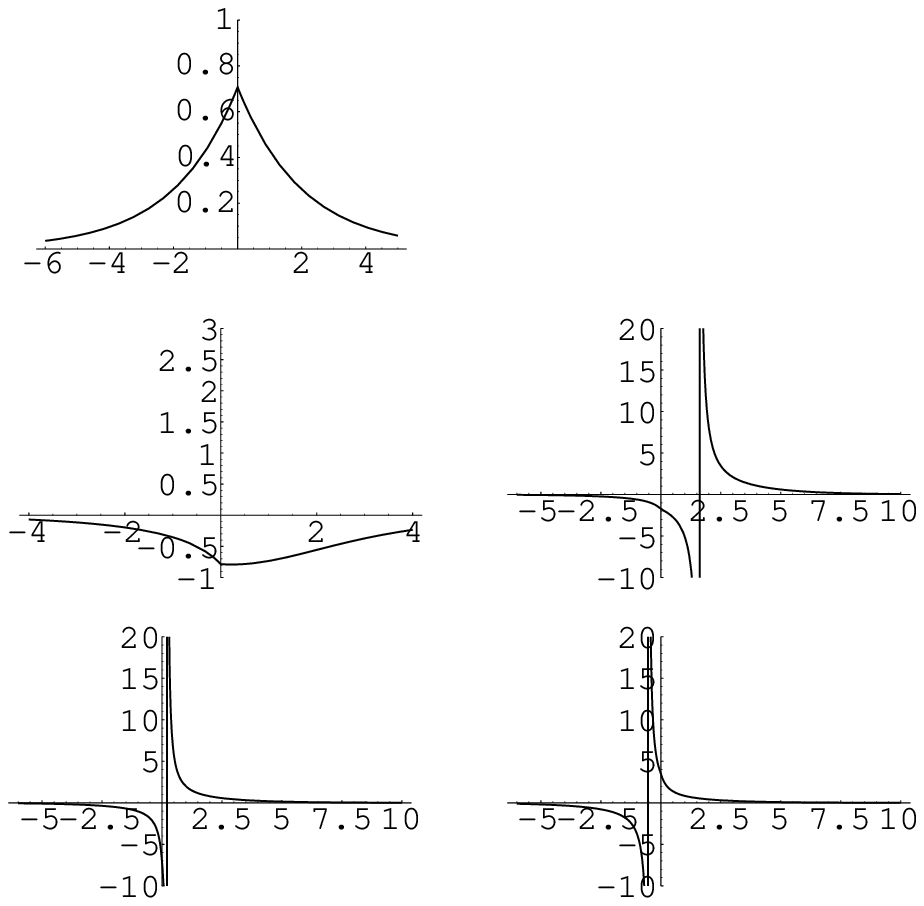}}
\vskip 4ex
\begin{center}
{\small{Figure 4}\\
}
\end{center}

\end{document}